\documentclass{article}
\usepackage{eurosym}
\usepackage{amsmath,amssymb,amsthm}
\usepackage{graphicx}
\usepackage{subcaption}
\usepackage{verbatim}
\usepackage{hyperref}
\usepackage{bm}
\usepackage{color}
\usepackage[numbers]{natbib}
\usepackage{amsmath}
\usepackage{amsfonts}
\usepackage{amssymb}

\begin{document}

\title{The Freedericksz transition in a spatially varying magnetic field}
\author{Tianyi Guo \\
%EndAName
Advanced Materials and Liquid Crystal Institute, Kent State University \and %
Xiaoyu Zheng \\
%EndAName
Department of Mathematica Sciences, Kent State University \and Peter
Palffy-Muhoray \\
%EndAName
Advanced Materials and Liquid Crystal Institute, and \\
\ Department of Mathematical\ Sciences, Kent State University}
\maketitle
\date{}

\begin{abstract}
Much is known about the Freedericksz transition induced by uniform electric
and magnetic fields in nematic liquid crystals. Here we study the effects of
a spatially varying field on the transition. We study the response of a
nematic to a magnetic field with cylindrical symmetry, and find that since
the field magnitude varies in the plane of the cell, the transition vanishes.
\end{abstract}

\section{Introduction}

Nematic liquid crystals are anisotropic fluids whose constituent particles
prefer to align, for reasons of energy and/or entropy, with their symmetry
axes parallel. This direction is typically designated by a unit vector $%
\mathbf{\hat{n}}$, the nematic director. In the absence of interactions with
surfaces or external fields, there is no preferred direction for $\mathbf{%
\hat{n}}$; all directions are equally likely. In 1913, Charles Mauguin
showed that the nematic director can be aligned by surfaces as well as a
magnetic field \cite{Mauguin, Maugin2} and in 1927, Vsevolod Freedericksz
showed that the nematic director can be aligned by an electric field \cite%
{Freedericksz}. If both surface and field alignment interactions are present
with different preferred directions, then a competition occurs; as the
applied field magnitude is increased, a transition occurs from the surface
dominated alignment to the field dominated alignment. This transition is the
Freedericksz transition.

One classical example of the Freedericksz transition occurs in a cell with
parallel plates with strong homeotropic alignment; that is, where the
surface alignment is such that the director $\mathbf{\hat{n}}$ is fixed
parallel to the surface normal. If a magnetic field $\mathbf{H}$ is applied
parallel to the plates, and if the symmetry axes of the molecules prefer to
align with the magnetic field in the absence of surface interactions, there
will be a competition between the surface anchoring and magnetic field
alignment effects. If the field $\mathbf{H}$ is sufficiently strong, the
director will change its direction and deviate from the surface normal to
align with the field. To do this, it must overcome the Frank \cite{Frank}
elastic energy associated with the spatial variation of the director. The
competition between the elastic energy and the interaction energy of the
liquid crystal with the field leads to the Freedericksz transition.

\begin{figure}
		\center
	\includegraphics[width=10.5 cm]{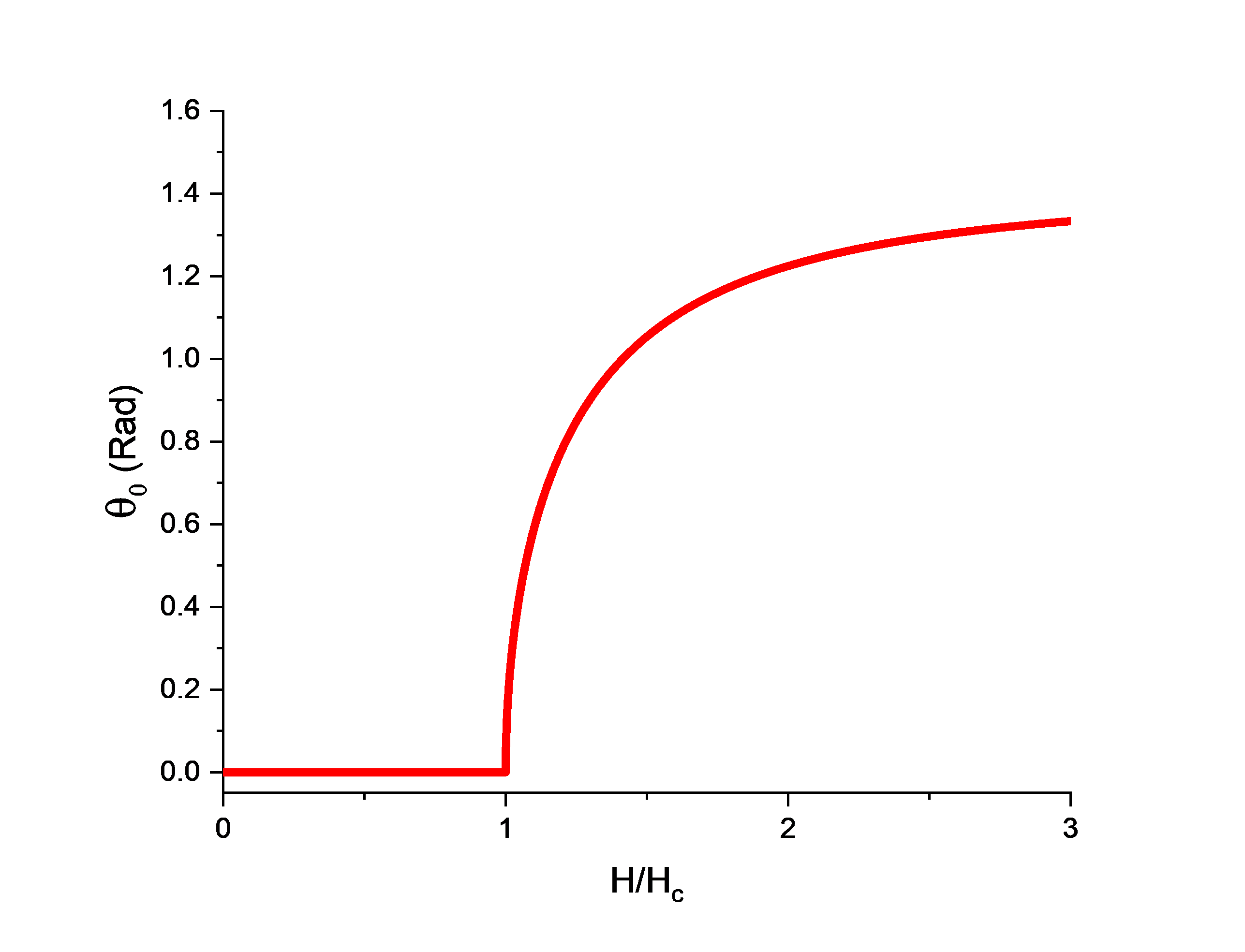}
	\caption{Deviation
			angle in the middle of the cell as function of $H/H_{c}$ in the classical
			Freedericksz transition. \label{fig1}}
\end{figure}

The dimensionless free energy density of the system can be written in terms
of the deviation angle $\theta _{0}$ in the middle of the cell, to leading
order, as \cite{Pieranski, de Gennes},%
\begin{equation}
\mathcal{F=}\frac{1}{2}\theta _{0}^{2}(1-\frac{H^{2}}{H_{c}^{2}})+\frac{1}{8}%
\frac{H^{2}}{H_{c}^{2}}\theta _{0}^{4},  \label{e1}
\end{equation}%
where $H_{c}$, representing both elastic and magnetic contributions, is the
critical magnetic field. Minimizing the free energy with respect to $\theta
_{0}$ give two solutions: 
\begin{equation}
\theta _{0}=0,
\end{equation}%
and%
\begin{equation}
\theta _{0}=\pm \sqrt{2(1-\frac{H_{c}^{2}}{H^{2}})},  \label{dis}
\end{equation}%
as shown in Fig.~\ref{fig1}.

For $H<H_{c}$, the only solution is $\theta _{0}=0$; that is to say, $%
\mathbf{\hat{H}}\cdot \mathbf{\hat{n}}=0$ for all $H<H_{c}$. As the field is
increased, at $H=H_{c}$ a second solution with $\theta _{0}\neq 0$ appears,
and for $H>H_{c}$, $\theta _{0}$ rapidly increases with $H$ and the solution
minimizes the free energy. The transition is second order, in accordance
with the Landau theory of second order phase transitions \cite{Landau}. The
classical Freedericksz transition, as described above has received
considerable attention. Since it relates elastic properties of liquid
crystals to susceptibilities to applied fields, it has been widely used to
proble the properties of nematics.

It is interesting to ask what happens if the magnetic field $H$ is not
uniform. If $H$ were to vary linearly with position $x$, $\theta _{0}$ might
be expected to depend on position as it does on the field $H$ in Fig. 1.
However, this would result in a large elastic energy cost due to the rapid
variation of $\theta _{0}$ with position.

The work reported here was carried out to answer this question, to find out
what happens in the Freedericksz transition when the field driving the
transition is not uniform, but varies in space. The details of this work are
given below.

\section{Experiment}

The experiment consisted of optical interferometric measurements on a
homeotropically aligned nematic liquid crystal in a planar circular cell
placed in a radial magnetic field. The magnitude of the field increased
monotonically from zero at the center to a maximum value at the cell
perimeter. This geometry provides a well characterized spatially varying
field with high symmetry, allowing study of the response of the nematic
liquid crystal to the spatially varying magnetic field. Details of the
experiment are provided below.

\subsubsection{Ring magnet}

The source of the magnetic field was a ring magnet consisting of twelve
wedge-shaped N50 rare earth neodymium magnets with dimensions ($49.95mm$ OD $%
\times \ 26mm$ ID $\times \ 11.5mm$ thickness) purchased from
Supermagnetman.net. The magnet generates a very nearly radially symmetric
magnetic field, pointing from the edge towards the center. In the mid-plane
of the ring magnet, the magnitude of the field varies along the radial
direction decreasing to zero at the center of the ring. The magnet, with its
axis horizontal, was mounted on an xyz stage, as shown in Fig.~\ref%
{setup_ring}. It can be translated in the z-direction to apply the magnetic
field to the coaxial sample cell shown on the right.

\begin{figure}
		\center
	\includegraphics[width=10.5 cm]{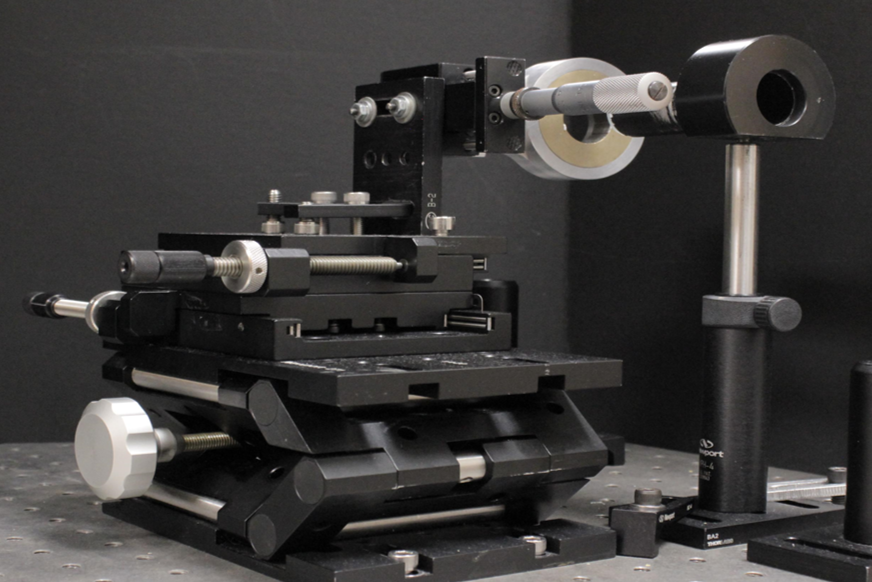}
	\caption{Ring magnet on xyz stage. \label{setup_ring}}
\end{figure}

A LakeShore Model $450$
Gaussmeter with a Hall probe was used to measure the field strength of the
ring magnet. The probe was mounted on a translation stage; the field was
measured as a function of radial position.

\subsubsection{Circular liquid crystal cell}

The liquid crystal cell consisting of two circular borosilicate glass
substrates ($25mm$OD$\times 0.1cm$) separated by $50{\mu}m$ annular Mylar film spacers and held together with Norland 65 adhesive was
designed to fit into the central cavity of the ring magnet. The cell was
filled 4-cyano-4 -pentylbiphenyl (5CB) via capillary action. Strong
homeotropic surface alignment was obtained using silane \cite{thesis}.

\subsubsection{Optical setup}

The nematic director in the magnetic field was studied using optical
interferometry. A spatial filter was used to obtain a clean, well aligned
beam from a Melles Griot $5mW$ laser at $543.5nm$ along the symmetry axis of
the cell. The beam was collimated to the diameter of the cell. The cell and
the magnet were placed between two crossed polarizers. Interference patterns
were recorded as the magnet was translated along the beam. The sample was at
room temperature. The position of the magnet relative to the cell was
measured and recorded manually while the interference pattern was displayed
on a screen and captured using a Canon EOS Rebel T2i camera.

\subsection{Theory}

\subsubsection{Modeling magnetic field}

\begin{figure}
		\center
	\includegraphics[width=10.5 cm]{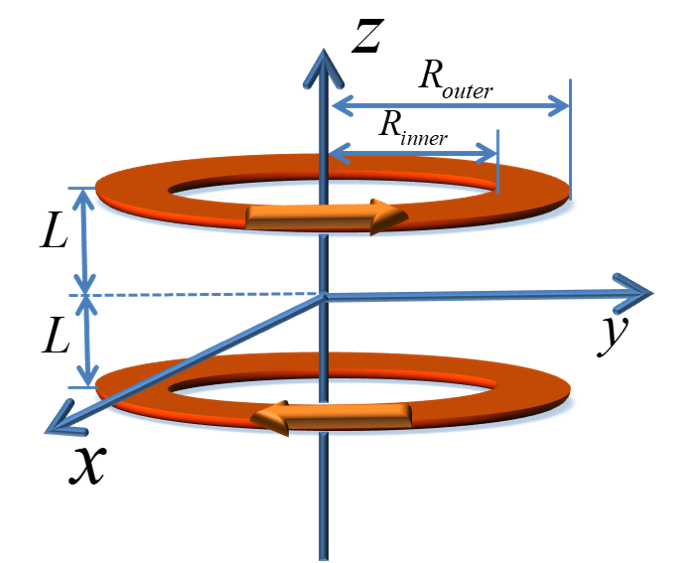}
	\caption{Two annuli
		with currents flowing in opposite directions on top and bottom of the ring
		magnet with separation of $2L$. \label{annuli}}
\end{figure}

The radially symmetric magnetic field was generated by a ring formed from
truncated wedge-shaped rare-earth magnets. Each magnet is viewed as a
surface current loop. Due to cancellation of surface currents at the contact
planes, there are two current annuli flowing in opposite directions
(counter-clockwise and clockwise) on top and bottom of the ring magnet as
shown in Fig.~\ref{annuli}.

The magnetic field is calculated for each current loop using numerical
integration of the Biot-Savart Law,%
\begin{equation}
\mathbf{B}=\frac{\mu _{0}}{4\pi }J\int_{R_{inner}}^{R_{outer}}\int_{0}^{2\pi
}\frac{\mathbf{\hat{J}}\times \mathbf{\hat{r}}}{r^{2}}Rd\phi dR,
\end{equation}%
where $\mathbf{J}$ is the surface current density, $\mathbf{r}$ is the
vector from the current element to the point where $\mathbf{B\,\ }$is
specified and $\phi $ is the azimuthal angle.

\subsubsection{Modeling and numerics for director field}

The director field is 
\begin{equation}
\mathbf{\hat{n}}=\left( \sin \theta ,0,\cos \theta \right),
\end{equation}%
where $\theta =\theta \left( r,z\right) $ is the angle between the director
and the glass substrate normal.

The free energy is%
\begin{equation}
F_{total}=\int\nolimits_{V}\left( \mathcal{F}_{elastic}+\mathcal{F}%
_{mag}\right) dV,
\end{equation}%
and in the one constant approximation, the elastic energy density is given by%
\begin{equation}
\mathcal{F}_{elastic}=\frac{1}{2}K\left( \frac{1}{r^{2}}\sin ^{2}\theta
+\left( \frac{\partial \theta }{\partial z}\right) ^{2}+\left( \frac{%
\partial \theta }{\partial r}\right) ^{2}-\frac{2}{r}\frac{\partial \theta }{%
\partial z}\sin ^{2}\theta +\frac{2}{r}\frac{\partial \theta }{\partial r}%
\sin \theta \cos \theta \right) ,
\end{equation}%
and when the cell is in the midplane, $\mathbf{H}=-H_{r}\mathbf{\hat{r}}$ and%
\begin{equation}
\mathcal{F}_{mag}=-\frac{1}{2}\mu _{o}\Delta \chi H_{r}^{2}\sin ^{2}\theta ,
\end{equation}%
where $\mu _{0}$ is the permeability of free space, and $\Delta \chi $ is
the diamagnetic susceptibility anisotropy. Since the system has cylindrical
symmetry and the cell is thin, the variation of $H$ in the $z-$ direction
was neglected.

Setting the dissipation rate equal to the rate of change of free energy
gives the dynamics for the approach to equilibrium. This gives the equation
of motion%
\begin{equation}
\gamma \dot{\theta}=\frac{\partial ^{2}\theta }{\partial z^{2}}+\frac{%
\partial ^{2}\theta }{\partial r^{2}}+\frac{1}{r}\frac{\partial \theta }{%
\partial r}-\frac{1}{2r^{2}}\sin 2\theta +\frac{1}{2}\alpha _{r}^{2}\sin
2\theta ,  \label{dy}
\end{equation}%
where $\gamma $ is a viscosity and%
\begin{equation}
\alpha _{r}=\sqrt{\frac{\mu _{o}\Delta \chi }{K}}H_{r}=\frac{\pi }{d}\frac{%
H_{r}}{H_{c}},
\end{equation}%
where $K$ is the elastic constant. Boundary conditions are $\theta
(0,z)=\theta (R,z)=0$ and $\theta (r,d/2)=\theta (r,-d/2)=0$.

Equation \eqref{dy} was solved numerically using a finite difference and
forward time stepping method \cite{thesis}. A $5\times 200$ square $z-r$
grid was used in the discretization. Initial conditions were biased random
orientation of the director on the lattice sites to prevent domain formation
in the sample. The numerics was implemented in Visual Fortran, and starting
from initial conditions, the code was executed until the mean squared change
in $\theta $ was below the threshold in the range of $10^{-14}-10^{-16}$.

\subsubsection{Modeling light propagation}

In the experiment, polarized monochromatic light is normally incident on the
sample and propagates along the symmetry axis. The sample is anisotropic and
inhomogeneous, so analytic solutions are not available, and the
inhomogeneity makes numerics challenging. Approximations are therefore made
in the model to make the problem tractable.

To calculate the transmitted intensity, it is convenient to work with the
electric displacement $\mathbf{D}$ instead of the electric field, since for
a plane wave with wave vector $\mathbf{k}=(2\pi /\lambda )\mathbf{\hat{z}}$, 
$\mathbf{k\cdot D}=0$. Then, for the incident light polarized along the $%
\mathbf{\hat{x}-}$ axis in the lab frame, the incident displacement is
given, in terms of the two normal modes, by 
\begin{equation}
\mathbf{D}_{i0}=D_{i0}(\mathbf{\hat{x}}\cdot \mathbf{\hat{r})\hat{r}+}D_{i0}(%
\mathbf{\hat{x}}\cdot \mathbf{\hat{\phi})\hat{\phi},}  \label{d}
\end{equation}%
where $\phi $ is the azimuthal angle such that $r\cos \phi =x$, and $\mathbf{%
\hat{\phi}=\hat{z}\times \hat{r}}$. The cell is regarded as being composed
of thin parallel nematic layers, with thickness $\Delta $ and layer normal $%
\mathbf{\hat{z}}$. The director is in the $z-r$ plane, and $\mathbf{\hat{n}%
\cdot z=}\cos \theta $. The refractive index for the mode in the $\mathbf{%
\hat{\phi}}$ direction is%
\begin{equation}
n_{1}=n_{o},
\end{equation}%
and for the mode in the $\mathbf{\hat{r}}$ direction, it is 
\begin{equation}
n_{2}=\frac{n_{e}n_{o}}{\sqrt{n_{e}^{2}\cos ^{2}\theta +n_{o}^{2}\sin
^{2}\theta }},
\end{equation}%
where $n_{e}$ and $n_{o}$ are the principal refractive indices of the liquid
crystal. The sample is assumed to be non-absorbing in the visible, and hence
the effect of propagation through a layer is only the acquisition of phase.
Reflections have been ignored. The ordinary mode along $\mathbf{\hat{\phi}}$
acquires phase%
\begin{equation}
\delta _{1}=\frac{2\pi n_{o}}{\lambda _{0}}d,
\end{equation}%
while the extraordinary mode along $\mathbf{\hat{r}}$ acquires phase%
\begin{equation}
\delta _{2}=\frac{2\pi n_{o}}{\lambda _{0}}d\sum_{j}\frac{n_{e}}{\sqrt{%
n_{e}^{2}\cos ^{2}\theta _{j}+n_{o}^{2}\sin ^{2}\theta _{j}}}\frac{\Delta }{d%
},
\end{equation}%
where $\lambda _{0}$ is the free space wavelength and $j$ refers to the $%
j^{th}$ nematic layer.

The light transmitted through the cell passes through the analyzer in the $%
\mathbf{\hat{y}}$ direction, and the transmitted displacement is%
\begin{equation}
\mathbf{D}_{t}=D_{t0}(e^{i\delta _{2}}(\mathbf{\hat{x}}\cdot \mathbf{\hat{r}%
)(\hat{y}\cdot \hat{r})+}e^{i\delta _{1}}(\mathbf{\hat{x}}\cdot \mathbf{\hat{%
\phi})(\hat{y}\cdot \hat{\phi}))\hat{y}.}
\end{equation}%
Since the intensity is proportional to $\mathbf{DD}^{\ast }$, it is given by 
\begin{equation}
I=I_{0}\sin ^{2}(\frac{\delta _{2}-\delta _{1}}{2})\sin ^{2}2\phi .
\end{equation}%
At given point $(x,y)$, $\phi $ satisfies%
\begin{equation}
\sin 2\phi =\frac{2xy}{x^{2}+y^{2}}.
\end{equation}%
Since $\theta (r)$ is known from simulations, the refractive indices $n_{1}$
and $n_{2}$ and the phase shifts $\delta _{1}$ and $\delta _{2}$ can be
determined, and the transmitted intensity at each point can be calculated.

\section{Results}

\subsection{Experiment and numerics}

\subsubsection{B field measurements}

\begin{figure}
		\center
	\includegraphics[width=14 cm]{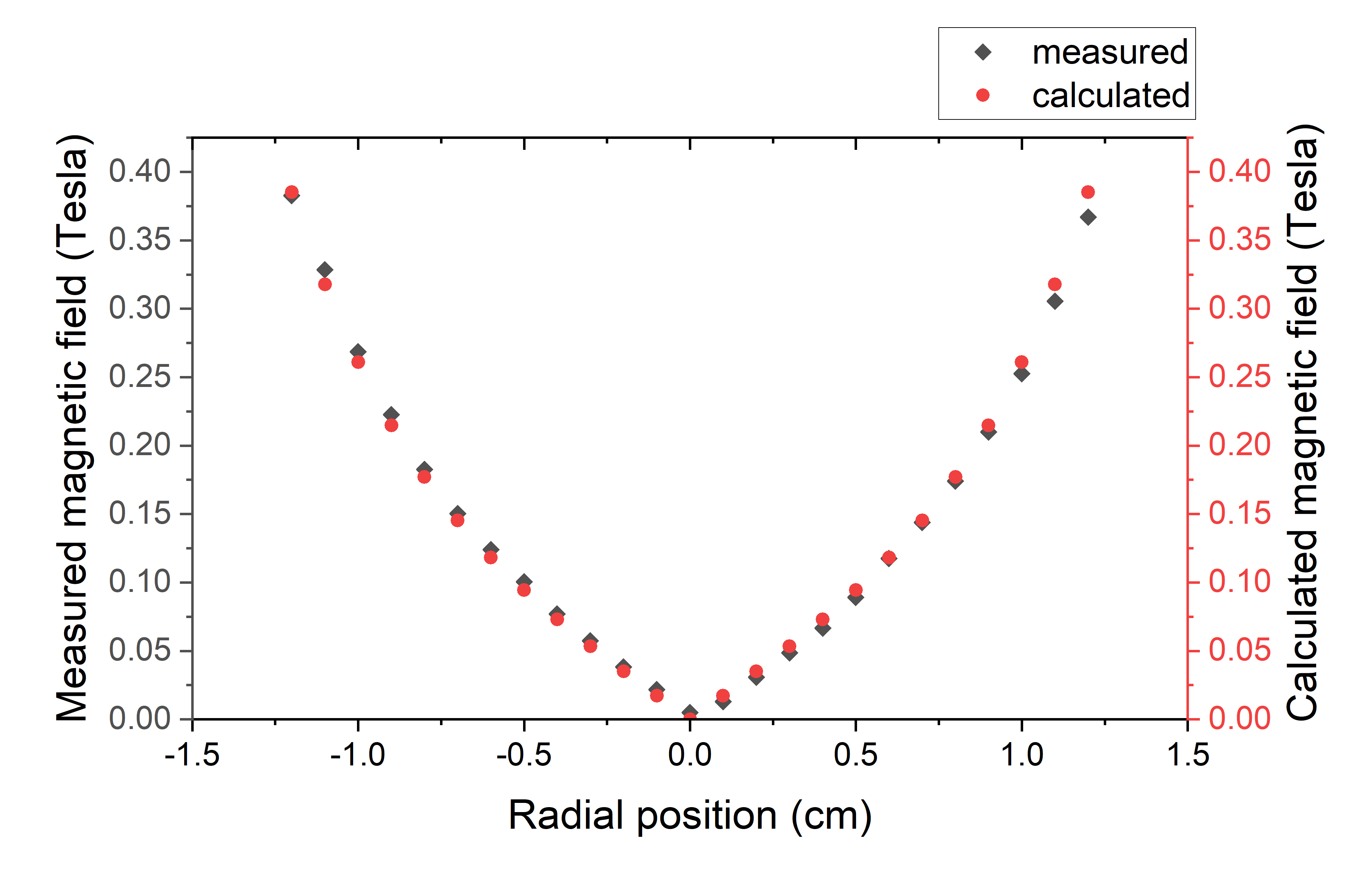}
	\caption{Calculated and measured magnetic field on the mid-plane as a function of radial position.\label{cal_magfield}}
\end{figure}

The calculated and measured magnetic fields in the mid-plane as a function
of radial position are shown in Fig.~\ref{cal_magfield}; they
agree to with $6\%$. The field is zero at the center due to symmetry and
increases initially linearly with radius $r$.

\subsubsection{Interference patterns}

Observed and calculated interference patterns for different cell positions,
relative to the ring magnet, in the $z-$direction, are shown in Fig.~ \ref%
{pattern_com}.

\begin{figure}
		\center
	\includegraphics[width=16 cm]{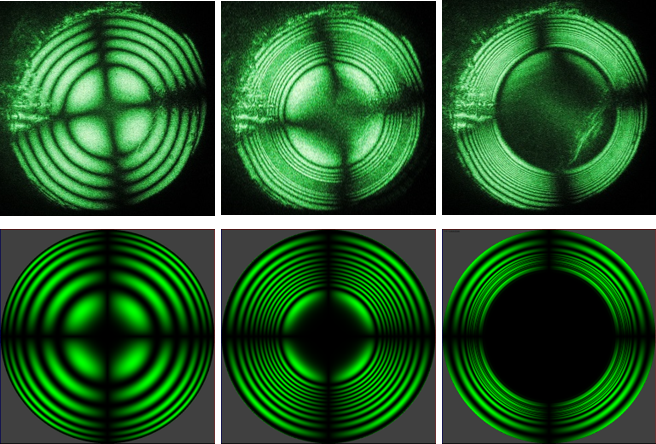}
	\caption{ Top: the experimental results of the far field interference
			patterns as the magnet is translated along the beam. Bottom: the simulation
			results of the patterns at the corresponding positions.\label{pattern_com}}
	\end{figure}

Magnified far field interference patterns on a screen were observed and
photographed (top row) in Fig.~\ref{pattern_com} as the magnet was
translated along the symmetry axis. On either side of the midplane, the
magnetic field has non-zero $z$ components, as does the director field. The
pattern on the left was captured when the sample was outside the ring, the
pattern in the middle is when the sample is in the ring, but not yet at the
midplane, and the pattern on the right is when the sample is in the
midplane. The corresponding patterns on the bottom row in Fig.~\ref%
{pattern_com} are calculated, based on the director field configuration
obtained from energy minimization \cite{thesis, TG mclc}.

\begin{figure}
			\center
		\includegraphics[width=6 cm]{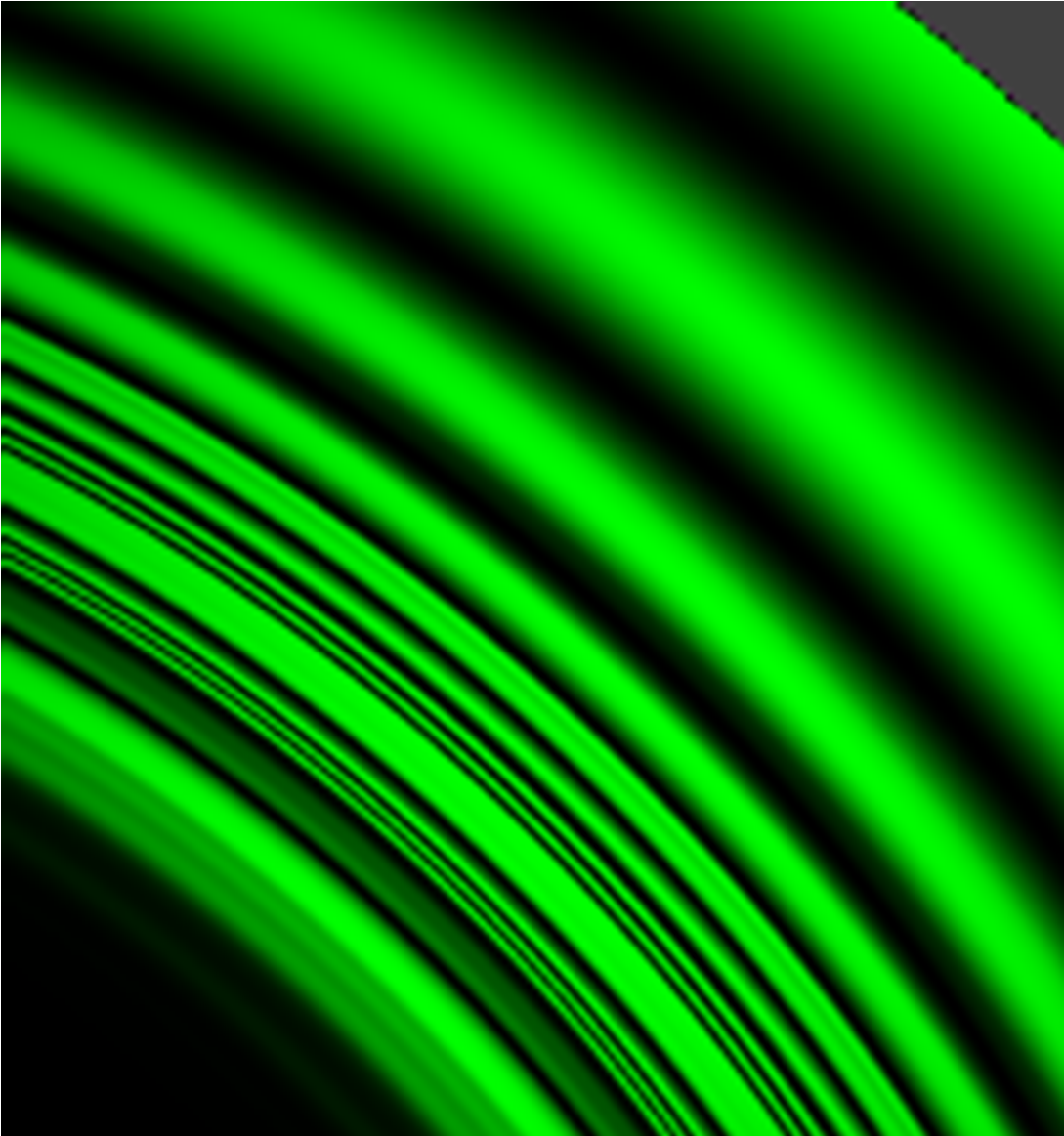}
		\caption{Enlarged portion of the bottom
			right pattern in Fig.~\ref{pattern_com}. \label{gen}}
	\end{figure}

	An enlarged version of the computer generated interference pattern is shown
	in Fig.~\ref{gen} to indicate details of the structure.
	
\subsubsection{Director field}

\begin{figure}
		\center
	\includegraphics[width=14 cm]{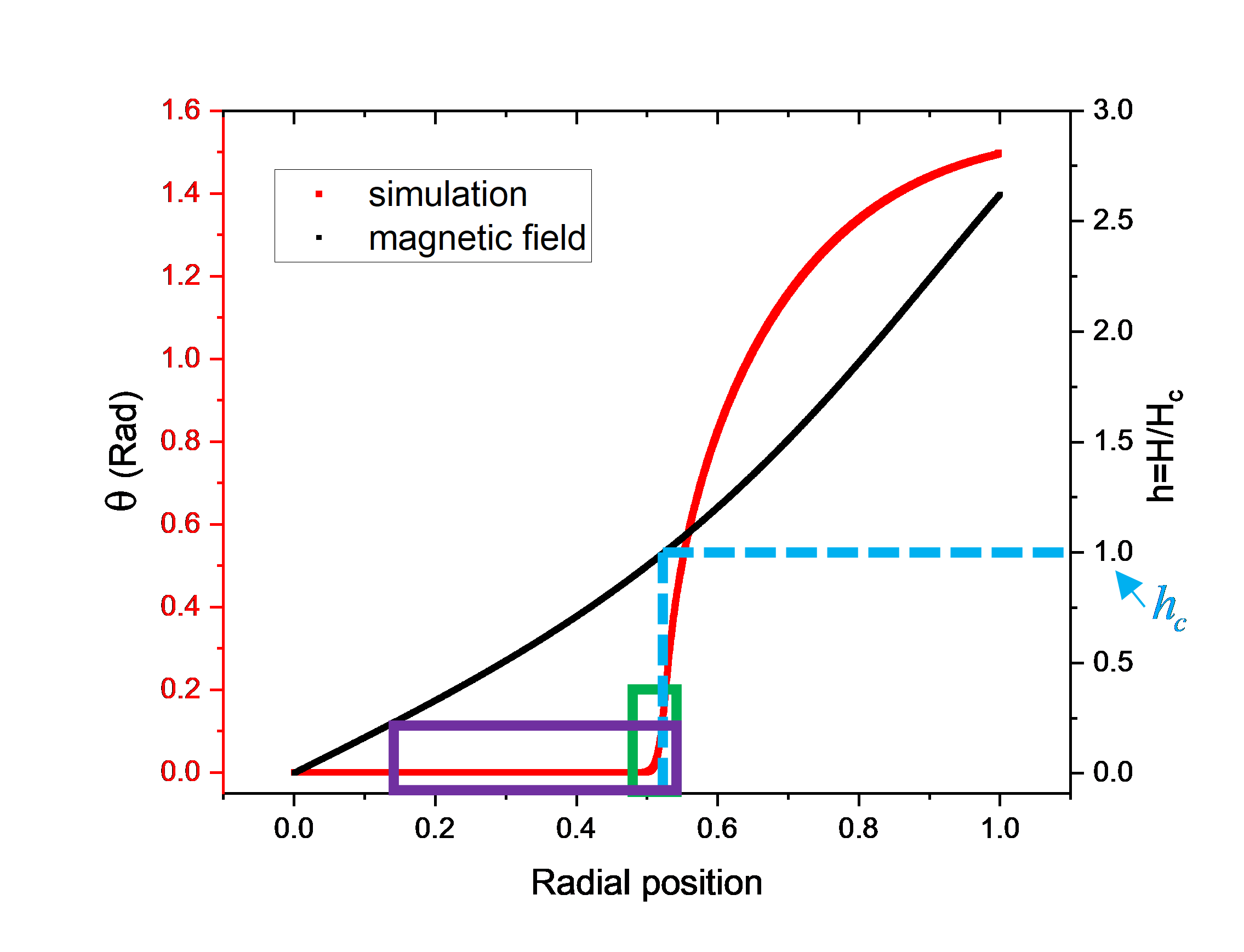}
	\caption{Director deviation angle $\theta _{0}$ and magnetic field $H/H_{c}$ as function of position $r$.\label{main}}
\end{figure}

\begin{figure}
		\center
	\includegraphics[width=16 cm]{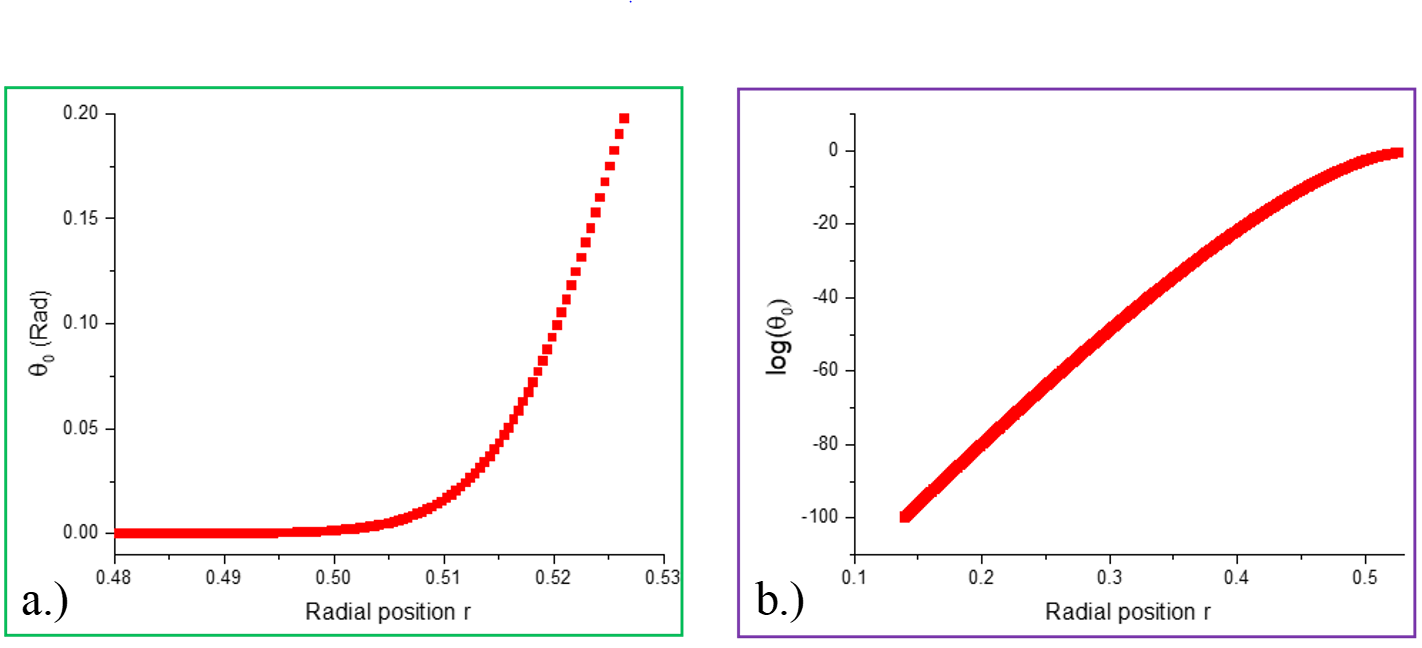}
	\caption{Details of Fig.~\ref{main}
		; a.) shows the decrease of $\theta _{0}$ just below the classical
		transition value $H_{c}$ as the magnetic field is reduced, in the box
		indicated, while b.) shows $\log \protect\theta _{0}$ vs.\ position $r$
		illustrating exponential-like decay, in the box indicated.\label{magn}}
\end{figure}

Determination of the director pattern in the spatially varying field is the
central aspect of this work. With the sample at the midplane of the ring
magnet, Fig.~\ref{main} shows the director angle $\theta _{0}$ in the
midplane of the sample, half-way between the plates. The magnetic field is
radial, increasing monotonically with $r$ as shown. The angle $\theta _{0}$
changes rapidly in the vicinity of $r=0.5$ where the field $H\approx
0.9H_{c} $ is below the critical value. Although superficially the plot of $%
\theta _{0}$ vs.\ $r$ is similar to Fig. 1, as the magnified portions in
Fig.~\ref{magn} indicate, instead of falling abruptly to zero, $\theta _{0}$
decreases more slowly, the curvature of the curve changes sign as seen
clearly in Fig.\ \ref{magn}b, and although $\theta _{0}$ is small, it is
non-zero, decreasing apparently exponentially as shown in Fig.~ \ref{magn}a
towards the origin.

In Fig.\ \ref{pattern_com}, the experimental top right figure suggests light
transmission - hence non-zero $\theta _{0}$- near the inner boundary of the
central dark region, becoming fainter towards the center, in qualitative
agreement with Fig.\ \ref{magn}b.

\subsection{Discussion}

We begin by noting that the region of interest in our cell is the vicinity
of the point where $H=H_{c}$, which is near $r=5mm$. We believe that in this
region, in our thin cell with $d=50\mu m$ , the difference between the
cylindrical coordinate and a Cartesian one is negligible, hence we disregard
the effects of cylindrical geometry.

Our results indicate that in our cell with strong homeotropic orientation,
the director remained partially aligned with the radial magnetic field even
in those regions of space where the magnetic field amplitude was below the
critical value $H_{c}$. That is to say, unlike in the case of the classical
Freedericksz transition, here $\mathbf{\hat{H}}\cdot \mathbf{\hat{n}}\neq 0$
for all $H<H_{c}$; the spatial variation of the field amplitude has
therefore fundamentally changed the nature of the transition. The reason for
the change can be understood from elementary considerations. At large values
of $r$, near the outer edge of the cell, the field is strong; $H\gg H_{c}$
and the director field in the midplane of the cell is expected to be well
aligned with the field, as in the classical case in Eq.~\eqref{e1}. Nearer
the center, the field magnitude is reduced, and hence the alignment will
also be reduced, as in the classical case. In the region where $H$ is just
above $H_{c}$, however, $\theta _{0}$ cannot depend on the field as in Eq.~\eqref{e1}; if it did, the gradient of the director field would diverge with
diverging elastic energy density. So instead of abruptly decreasing to zero, 
$\theta _{0}$ must decrease gradually, as shown in Fig.~\ref{magn}b.
Numerics indicates that the decrease remains gradual and exponential like, as
suggested by Fig.~\ref{magn}a.

\begin{figure}
	\center
	\includegraphics[width=10.5 cm]{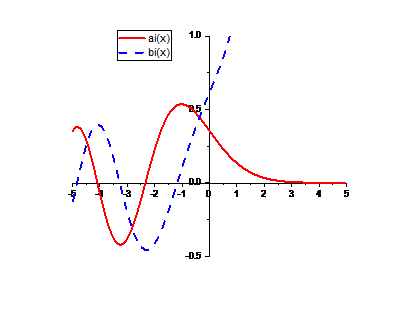}
	\caption{The airy functions $ai(x)$ and $bi(x)$.
		\label{airy}}
\end{figure}

More formally, since the field $H$ depends on position, $\theta _{0}$ will
depend on position, and if $\theta (r,z)=\theta _{0}(r)\cos (qz)$, the
dimensionless free energy density becomes, for small $\theta _{0}$, after
integrating over $z,$ 
\begin{equation}
\mathcal{F}=\frac{1}{2}\theta _{0}^{2}(1-\frac{H^{2}}{H_{c}^{2}})+\frac{d^{2}%
}{\pi ^{2}}(\frac{\partial \theta _{0}}{\partial r})^{2}.  \label{free}
\end{equation}%
To describe the spatial variation of $H$, we define $x=(r-r_{c})/r_{c}$, so
that $x=0$ is the location in the cell where $H=H_{c}$. In this vicinity, $x$
is small, and then, to a good approximation,%
\begin{equation}
H^{2}=H_{c}^{2}(1+x).
\end{equation}%
The Euler-Lagrange equation minimizing the free energy in Eq.~ \eqref{free}
then becomes%
\begin{equation}
\frac{\partial \theta _{0}^{2}}{\partial x^{2}}=-(\pi \frac{r_{c}}{d}%
)^{2}\theta _{0}x.
\end{equation}%
The solutions are Airy functions, shown in Fig.~\ref{airy}.

Since $ai$ dominates,%
\begin{equation}
\theta _{0}(x)\simeq ai(-\gamma x),
\end{equation}%
where%
\begin{equation}
\gamma =(\pi \frac{r_{c}}{d})^{\frac{2}{3}}.
\end{equation}%
Since $r_{c}=0.52$ corresponds to $6.6mm$, $\gamma =56$. We note that $%
ai(\gamma (r_{c}-r)/r_{c})$ is in good agreement with the gradual decrease
of $\theta _{0}(r)$ with decreasing $r$ near $r_{c}$ as seen in Fig. \ref%
{magn}b, and the asymptotics for large $-x$ is exponential-like decay as
seen in Fig.~\ref{magn}a. Although it appears that $\theta _{0}$ decays to
zero at the origin (but not before), this has not been confirmed. As Fig.~\ref{magn}a incidates, the values of $\theta _{0}$ near the origin are
extremely small, and in the numerical computation of the director field, the
relaxation of Eq.~\eqref{dy} is extremely slow. We are not yet therefore
certain, but the indications are that $\theta _{0}$ indeed decays to zero at
the origin. The spatial variation of the magnetic field therefore `lifts'
the system from the transition, and if the magnetic field is not
homogeneous, there are no discontinuities either in the director field, or
in its derivatives. This is reminiscent of the effect of an electric or
magnetic field on the temperature driven nematic-isotropic phase transition;
the presence of the field makes the order parameter everywhere nonzero, and
it and its derivatives continuous.

\subsection{Conclusions}

The response of a nematic cell in a spatially varying magnetic field was
explored experimentally and theoretically. The magnetic field with
cylindrical symmetry was generated by a series of wedge-shaped rare earth
magnets forming a ring. The strength of the field in the radial direction in
the midplane of the ring magnet was measured and calculated. The director
configuration of a homeotropically aligned nematic in this field was probed
using interferometry. The director field was modeled using the Oseen-Frank
theory, and determined numerically via an energy minimizing scheme.
Agreement of the calculated and measured interference patterns verified the
calculated director field. Of particular interest in this project was the
behavior of the nematic director in the region where the magnetic field
magnitude is near the critical field value in the classical Freedericksz
transition. The results indicate a smoothly varying director field
everywhere, with no evidence of discontinuities of the director angle or its
derivatives. Although the director field studied here differs only slightly
from what it would be in a uniform field with the same magnitude, the
director field changes smoothly with position everywhere in the regions
studied. The effect of the spatial variation of the magnetic field is the
absence of discontinuities; in a spatially varying field therefore there is
no Freedericksz transition. We anticipate equivalent result in planar
aligned rather than homeotropic cells, in the case of electric rather than
magnetic field induced transitions, in the case of wedge cells and in other
analogous system.

\section{Acknowledgments}

This work was supported by the Office of Naval Research through the MURI on
Photomechanical Material Systems (ONR N00014-18-1-2624).

\end{document}